\documentstyle[12pt,titlepage,epsbox]{article}
\begin{document}
\pagestyle{empty}
\baselineskip=0.212in

\begin{flushleft}
\large
{SAGA-HE-83-95, ECT$^\star$/NOV 95/03
\hfill November 30, 1995}  \\
%%%\hfill \today}  \\
\end{flushleft}

\vspace{1.0cm}

\begin{center}

\LARGE{{\bf  Nuclear dependence of $\bf Q^2$ evolution}} \\

\vspace{0.3cm}

\LARGE{{\bf  in the structure function $\bf F_2$ }} \\

\vspace{1.0cm}

\Large
{S. Kumano} \\

\vspace{0.2cm}

\Large
{Department of Physics, Saga University, Saga 840, Japan} \\

\vspace{0.2cm}

\Large
{and ECT$^\star$, Villa Tambosi, I-38050 Villazzano (Trento), Italy} \\

\vspace{0.5cm}

\Large
{and M. Miyama}         \\

\vspace{0.2cm}

\Large
{Department of Physics, Saga University, Saga 840, Japan $^\dagger$} \\

\vspace{1.0cm}

\Large{ABSTRACT}

\end{center}

$Q^2$ evolution of the structure functions $F_2$
in tin and carbon nuclei is investigated in order to
understand recent NMC measurements.
$F_2$ is evolved by using
leading-order DGLAP, next-to-leading-order DGLAP,
and parton-recombination equations.
NMC experimental result
$\partial [F_2^{Sn}/F_2^C]/ \partial [\ln Q^2]\ne 0$ could be
essentially understood by the difference of parton distributions
in the tin and carbon nuclei.
However, we find an interesting indication that large higher-twist
effects on the $Q^2$ evolution could be ruled out.
Nuclear dependence of the $Q^2$ evolution could be
interesting for further detailed studies.

\vspace{1.0cm}

\vfill

\noindent
{\rule{6.cm}{0.1mm}} \\

\vspace{-0.2cm}
\normalsize
\noindent
{$\dagger$ Email: kumanos or 94sm10@cc.saga-u.ac.jp.
   Information on their research is available}  \\

\vspace{-0.6cm}
\noindent
{at http://www.cc.saga-u.ac.jp/saga-u/riko/physics/quantum1/structure.html} \\

\vspace{-0.6cm}
\noindent
\normalsize
{or at ftp://ftp.cc.saga-u.ac.jp/pub/paper/riko/quantum1.} \\

\vspace{1.0cm}

\vspace{-0.5cm}
\hfill
{submitted for publication}

\vfill\eject
%%%%%%%%%%%%%%%%%%%%%%%%%%%%%%%%%%%%%%%%%%%%%%%%%%%%%%%%%%%%%%%%%%%%%%
%%%%%%%%%%%%%%%%%%%%%%%%%%%%%%%%%%%%%%%%%%%%%%%%%%%%%%%%%%%%%%%%%%%%%%
\pagestyle{plain}

\noindent
{\Large\bf {1. Introduction}}
\vspace{0.4cm}

Nuclear modification of the structure function $F_2$ has
been an interesting topic since the discovery of the EMC
(European Muon Collaboration) effect in 1983.
Although most studies discuss $x$ dependence of the modification,
$Q^2$ dependence becomes increasingly interesting.
It is because the NMC (New Muon Collaboration) showed
$Q^2$ variations of the ratio $F_2^A/F_2^D$ with reasonably
good accuracy \cite{NMCQ2}.
Structure functions $F_2$ themselves cannot be calculated
without using nonperturbative methods;
however, $Q^2$ evolution of $F_2$ can be evaluated perturbatively.
The phenomenon is called scaling violation, which
is considered to be a strong evidence to support perturbative QCD.
An intuitive way of describing the scaling violation is
to use the Dokshitzer-Gribov-Lipatov-Altarelli-Parisi (DGLAP)
equations \cite{DGLAP}.
$Q^2$ variations of $F_2^A/F_2^D$
are calculated by the DGLAP equations \cite{SKF2,MK},
and they are consistent with existing NMC experimental data.

It is not, however, obvious whether DGLAP could be applied
to the nuclear case. Complex nuclear interactions
may give rise to extra $Q^2$ factors in the evolution equations.
In particular, the longitudinal localization size of a parton
with momentum fraction $x$ could exceed an average nucleon
separation in a nucleus if $x$ is small ($x<0.1$).
In this case, partons in different nucleons could interact
and the interactions are called parton recombinations.
Their contributions to the evolution equations have
$A^{1/3}$ nuclear dependence \cite{MQ}.
Until recently, no nuclear dependence of the Q$^2$ evolution
had been found experimentally.
In fact, previous measurements
are consistent with no nuclear dependence
$\partial [F_2^{A}/F_2^D]/ \partial [\ln Q^2]= 0$ \cite{NMC95}.
However, it is reported recently by NMC \cite{NMCSN-C} that
there exist significant differences between tin and carbon
$Q^2$ evolutions,
$\partial [F_2^{Sn}/F_2^C]/ \partial [\ln Q^2]\ne 0$.
It is the first indication of nuclear effects
in the $Q^2$ evolution of $F_2$ and is worth investigating
theoretically.

The purpose of our study is to calculate the nuclear
dependence $\partial [F_2^{Sn}/F_2^C]/ \partial [\ln Q^2]$
theoretically and to compare the results with the NMC data.
There are two major sources for the difference in
$Q^2$ evolution of $F_2$.
First, parton distributions themselves are different
in tin and carbon nuclei.
Second, nuclear interactions could modify the evolution equations.
For example, the parton recombinations
produce extra $Q^2$ dependent effects on the $Q^2$
evolution equations. Numerical solutions of DGLAP equations
and those with recombination effects are discussed
in Ref. \cite{MK}.

In section 2, we employ input distributions obtained in a hybrid parton
model with rescaling and recombination effects \cite{SKF2},
then the $Q^2$ evolution is calculated by using the computer code
in Ref. \cite{MK}.
Three types of evolutions are tested, and they are
leading-order (LO) DGLAP equations, next-to-leading-order (NLO) DGLAP,
and evolution equations with parton recombinations (PR).

\vfill\eject
%%%%%%%%%%%%%%%%%%%%%%%%%%%%%%%%%%%%%%%%%%%%%%%%%%%%%%%%%%%%%%%%%%%%%%
\vspace{1.0cm}
\noindent
{\Large\bf {2. $Q^2$ evolution of $\bf F_2$ in tin and carbon nuclei}}
\vspace{0.4cm}

Scaling violation has been well studied, and
it corresponds to the physics that minute quark-gluon clouds
around a parton could be seen by increasing $Q^2$.
Evolution equations are derived by calculating
parton splitting processes into two partons or by calculating
corresponding anomalous dimensions.
Parton-recombination contributions can also be included
in the equations.
$Q^2$ evolution is then described by
integrodifferential equations, and
the DGLAP \cite{DGLAP} and PR \cite{MQ} evolution
equations are given by \cite{MK}
$$
{\partial \over {\partial t}} \ q_i \left({x,t}\right)\
=\ \int_{x}^{1}{dy \over y}\
\left[\ \sum_j P_{q_{i} q_{j}}\left({{x \over y}}\right)\
           q_j \left({y,t}\right)\
+\  P_{qg}\left({{x \over y}}\right)\
g\left({y,t}\right)\ \right]
$$
$$
\ \ \ \ \ \ \ \ \ \ \ \ \ \ \ \ \ \ \ \ \ \ \ \ \ \ \
\ \ \ \ \ \ \ \ \ \ \ \ \ \ \ \ \ \ \ \ \ \ \ \ \ \ \
+ \ \left( recombination\ terms\ \propto \
{{\alpha_s A^{1/3}} \over {Q^2}} \right) \
\ ,
\eqno{(1a)}
$$
$$
{\partial \over {\partial t}} \ g\left({x,t}\right)\
=\ \int_{x}^{1}{dy \over y}\
\left[\ \sum_j P_{gq_j}\left({{x \over y}}\right)\
q_j \left({y,t}\right)\
+\ P_{gg}\left({{x \over y}}\right)\
g\left({y,t}\right)\ \right]
$$
$$
\ \ \ \ \ \ \ \ \ \ \ \ \ \ \ \ \ \ \ \ \ \ \ \ \ \ \
\ \ \ \ \ \ \ \ \ \ \ \ \ \ \ \ \ \ \ \ \ \ \ \ \ \ \
+ \ \left( recombination\ terms\ \propto \
{{\alpha_s A^{1/3}} \over {Q^2}} \right) \
\ ,
\eqno{(1b)}
$$
where the variable $t$ is defined by
$t = -(2/\beta_0) \ln [\alpha_s(Q^2)/\alpha_s(Q_0^2)]$.
$\alpha_s$ is the running coupling constant and $\beta_0$
is given by $\beta_0=(11/3)C_G-(4/3)T_R N_f$ with
$C_G=N_c$ (number of color), $T_R=1/2$, and $N_f$=number of flavor.
In the PR evolution case, there is an extra evolution
equation for a higher-dimensional gluon distribution.
Explicit expressions of recombination contributions are found
in Ref. \cite{MK,MQ}.
The first two terms in Eqs. (1a) and (1b) describe the process
that a parton $p_j$ with the nucleon's momentum fraction $y$
splits into a parton $p_i$ with the momentum fraction $x$
and another parton.
The splitting function $P_{p_i p_j}(z)$ determines
the probability that such a splitting process occurs and
the $p_j$-parton momentum is reduced by the fraction $z$.

Because the splitting functions $P_{p_i p_j}$ are independent of
nuclear interactions, there are two possibilities for
the nuclear dependence in Eqs. (1a) and (1b).
First, parton $x$ distributions are modified in a nucleus.
It is known that quark distributions in a nucleus are
effectively reduced at small $x$ due to shadowing mechanisms
and medium $x$ due to binding, rescaling, and other effects.
They are enhanced at large $x$ due to nucleon Fermi motion effects
and also at $x \approx 0.1$.
Modification of the gluon distribution in a nucleus is not
well known at this stage. In particular, there is little information
from the experimental side. However, we expect that the gluon
distribution is shadowed at small $x$ \cite{SKGLUE}.
If the parton distributions are modified in Eqs. (1a) and (1b),
they give rise to different $Q^2$ evolution through the
splitting processes.
Therefore, the nuclear dependence comes entirely from
the parton-distribution differences in the DGLAP evolution case.

Second, the parton-recombination mechanism supplies additional
$\alpha_s A^{1/3}/Q^2$ effects in Eqs. (1a) and (1b).
As it is mentioned in the introduction, partons in different
nucleons cannot be considered independent at small $x$
because the parton localization size becomes larger than
the nucleon size. The recombination probability is proportional
to the number of nucleons in the longitudinal direction,
so that there exists the factor $A^{1/3}$.
The factor $\alpha_s/Q^2$ arises from the fact that
the parton-parton recombination cross section is
proportional to $\alpha_s/Q^2$.
Hence, nuclear dependence exists in the recombination
terms as well as in the parton distributions
if the PR evolution is used.

In order to calculate $Q^2$ evolution of nuclear $F_2$ structure
functions, we need to have input parton distributions at
certain $Q^2$. Because the distributions themselves cannot be
calculated exactly, they depend on a used model.
In the present investigation, we should employ a model which can
at least explain measured ratios $F_2^A/F_2^D$.
As such a parton-model candidate, we have a hybrid model with
parton-recombination and $Q^2$-rescaling mechanisms \cite{SKF2}.
According to this model, both parton recombinations and $Q^2$ rescaling
are calculated at $Q_0^2$=0.8 GeV$^2$. Then, parton distributions
are evolved to those at larger $Q^2$. The model can explain
measured $x$ and $Q^2$ dependence of the ratio $F_2^A/F_2^D$ by NMC.
For the details of the model, we refer the reader to Ref. \cite{SKF2}.
We evolve the initial distributions at $Q_0^2$ in the parton model
to those at larger $Q^2$ by using the LO-DGLAP, NLO-DGLAP, and PR
evolution equations with the help of the computer code in Ref. \cite{MK}.
We first checked $x$ dependence of the ratio $F_2^{Sn}/F_2^C$.
Evolved results at $Q^2$=5 GeV$^2$ are compared with
the $x$ dependent ratios measured by NMC.
We find that they essentially agree with each other; however,
the model slightly underestimates (overestimates)
the ratio at small (medium) $x$.

Calculated results for $\partial [F_2^{Sn}/F_2^C]/ \partial [\ln Q^2]$
are shown at $Q^2$=5 GeV$^2$ together with preliminary NMC data
\cite{NMCSN-C} in Fig. 1.
The dotted, solid, and dashed curves correspond to
LO-DGLAP, NLO-DGLAP, and PR evolution results respectively.
The QCD scale parameter is $\Lambda$=0.2 GeV and
the number of flavor is three.
The DGLAP evolution curves agree roughly with the experimental tendency;
however, they underestimate the $Q^2$ variation
in the region $0.01<x<0.05$.
In the PR evolution, NLO effects in the DGLAP part are included.
It is interesting to find that the PR results disagree with
experimental data even in the sign.
The large discrepancy from the DGLAP results
is caused partly by the evolution from small $Q^2$ (0.8 GeV$^2$)
to 5 GeV$^2$. The recombination contributions
are higher-twist effects, so that they are very large in the small
$Q^2$ region. Because of the significant discrepancy from the data,
large parton-recombination contributions could be ruled out.
However, it does not mean that the PR evolution itself is in danger.
There is an essential parameter $K_{HT}$, which determines how large
the higher-dimensional gluon distribution is
($xG_{HT}(x,Q_0^2)=K_{HT}[xg(x,Q_0^2)]^2$).
We chose $K_{HT}$=1.68 so that $G_{HT}$ contribution
is 10\% to the gluon distribution \cite{MQ}.
However, as it is discussed in Ref. \cite{MQ},
the magnitude of $K_{HT}$ is unknown at this stage.
In order to discuss the validity of the PR evolution,
the constant $K_{HT}$ must be evaluated theoretically.

Because average $Q^2$ values of the NMC data are different at
each $x$ point, we should take into account the $Q^2$ variation
of the derivative.
According to the NMC's preliminary results, the average $Q^2$ in the
$x$=0.01 region is about a few GeV$^2$ and it is about 10--20 GeV$^2$
in the $x$=0.1 region. In order to show theoretical $Q^2$ variations,
the derivative is calculated at $Q^2$=2, 5, 20 GeV$^2$
in the NLO-DGLAP case, and the results are shown in Fig. 2.
The dashed, solid, and dotted curves correspond to
the results at 2, 5, 20 GeV$^2$ respectively.
In the $x$=0.01 region, the theoretical results with $Q^2$=2 GeV$^2$
agree with the NMC data.
The results with $Q^2$=20 GeV$^2$ also agree roughly with data
in the $x$=0.1 region.
However, we still underestimate the derivative between these regions.

{}From these analyses, we find that the essential part of
the NMC results could be understood within the parton-model
framework of Ref. \cite{SKF2} together with the usual $Q^2$
evolution equations \cite{MK}.
However, there are still discrepancies between the theoretical
evolution results and the NMC data. So we check the sensitivity
of the theoretical results on sea-quark and gluon modifications
in nuclei.
If the parton distributions in the tin and carbon nuclei are
identical, the $Q^2$ derivative
$\partial [F_2^{Sn}/F_2^C]/ \partial [\ln Q^2]$ has to vanish
in the DGLAP evolution.
Therefore, the finite values in the DGLAP cases
reflect nuclear modification of quark and gluon distributions.
If the sea-quark shadowing is increased significantly
at small $x$ in the tin
with keeping same distributions in the carbon,
the theoretical results agree with the NMC data
in the region $x\approx 0.03$.
On the other hand, the gluon shadowing is increased in the similar way,
the disagreement becomes slightly larger.
The sea-quark modification has more significant effects
on the $Q^2$ derivative than the gluon modification does.
In order to compare theoretical results with the NMC data
more seriously, detailed studies on nuclear parton distributions
as well as on nuclear $Q^2$ evolution equations are necessary.

\vfill\eject
%%%%%%%%%%%%%%%%%%%%%%%%%%%%%%%%%%%%%%%%%%%%%%%%%%%%%%%%%%%%%%%%%%%%%%
\vspace{1.0cm}
\noindent
{\Large\bf {3. Conclusions}}
\vspace{0.4cm}

We find that the NMC finding of nuclear $Q^2$ dependence,
$\partial [F_2^{Sn}/F_2^C]/ \partial [\ln Q^2]\ne 0$,
could be essentially understood by ordinary $Q^2$ evolution
equations. The most important factor for the derivative
is the nuclear modification of parton distributions
(especially sea-quark distributions) and
is not the $Q^2$ evolution due to the parton recombinations.
Our parton model together with the evolution equations
describes the NMC data fairly well, although it
slightly underestimate the nuclear difference.
In our analysis, ``large" higher-twist effects from
the parton recombinations could be ruled out.
However, it is encouraging to study the details of the
recombination mechanism in comparison with the NMC data.
Furthermore, studies of $x$ dependence in nuclear parton distributions
are essential for understanding the details of the NMC data.

$~~~$

$~~~$

%%%%%%%%%%%%%%%%%%%%%%%%%%%%%%%%%%%%%%%%%%%%%%%%%%%%%%%%%%%%%%%%%%%%%%
%%%%%%%%%%%%%%%%%%%%%%%%%%%%%%%%%%%%%%%%%%%%%%%%%%%%%%%%%%%%%%%%%%%%%%
\noindent
{\Large\bf {Acknowledgments}}
\vspace{0.4cm}

S.K. and M.M. thank communications with A. M\"ucklich and
A. Sandacz on NMC experimental results.
This research was partly supported by the Grant-in-Aid for
Scientific Research from the Japanese Ministry of Education,
Science, and Culture under the contract number 06640406.
S.K. thanks the European Centre for Theoretical Studies
in Nuclear Physics and Related Areas (ECT$^\star$) in Trento
for its hospitality and for partial support for this project.
He also thanks the Japanese Ministry of Education for supporting
his travel to the ECT$^\star$.

\vfill\eject
%%%%%%%%%%%%%%%%%%%%%%%%%%%%%%%%%%%%%%%%%%%%%%%%%%%%%%%%%%%%%%%%%%%%%%
%%%%%%%%%%%%%%%%%%%%%%%%%%%%%%%%%%%%%%%%%%%%%%%%%%%%%%%%%%%%%%%%%%%%%%

%%\vspace{3.0cm}
%%%%%%%%%%%%%%%%%%%%%%%%%%%%%%%%%%%%%%%%%%%%%%%%%%%%%%%%%%%%%%%%%%%%%%%%%%%%%%

\vfill\eject
%%%%%%%%%%%%%%%%%%%%%%%%%%%%%%%%%%%%%%%%%%%%%%%%%%%%%%%%%%%%%%%%%%%%%%%%%%%%%%
\noindent
{\Large\bf{Figure Captions}} \\

\vspace{-0.38cm}
\begin{description}
   \item[Fig. 1]
Nuclear dependence of $Q^2$ evolution is calculated at $Q^2$=5 GeV$^2$.
The dotted, solid, and dashed curves are the results in the LO-DGLAP,
NLO-DGLAP, and PR evolutions respectively.
$K_{HT}$=1.68 is taken in the PR evolution \cite{MQ}.
The theoretical results are compared with the preliminary NMC data
\cite{NMCSN-C}.
   \item[Fig. 2]
The derivative is calculated at $Q^2$=2, 5, and 20 GeV$^2$
in order to show $Q^2$ variations.
The evolution method is the NLO-DGLAP.
The dashed, solid, and dotted curves correspond to
the results at $Q^2$=2, 5, and 20 GeV$^2$ respectively.
They are compared with the NMC data.
\end{description}

\vfill\eject
%%%%%%%%%%%%%%%%%%%%%%%%%%%  figure 1 %%%%%%%%%%%%%%%%%%%%%%%%%%%%%%%%%%
\vspace*{-5.0cm}
\hspace*{-1.2cm}
\epsfile{file=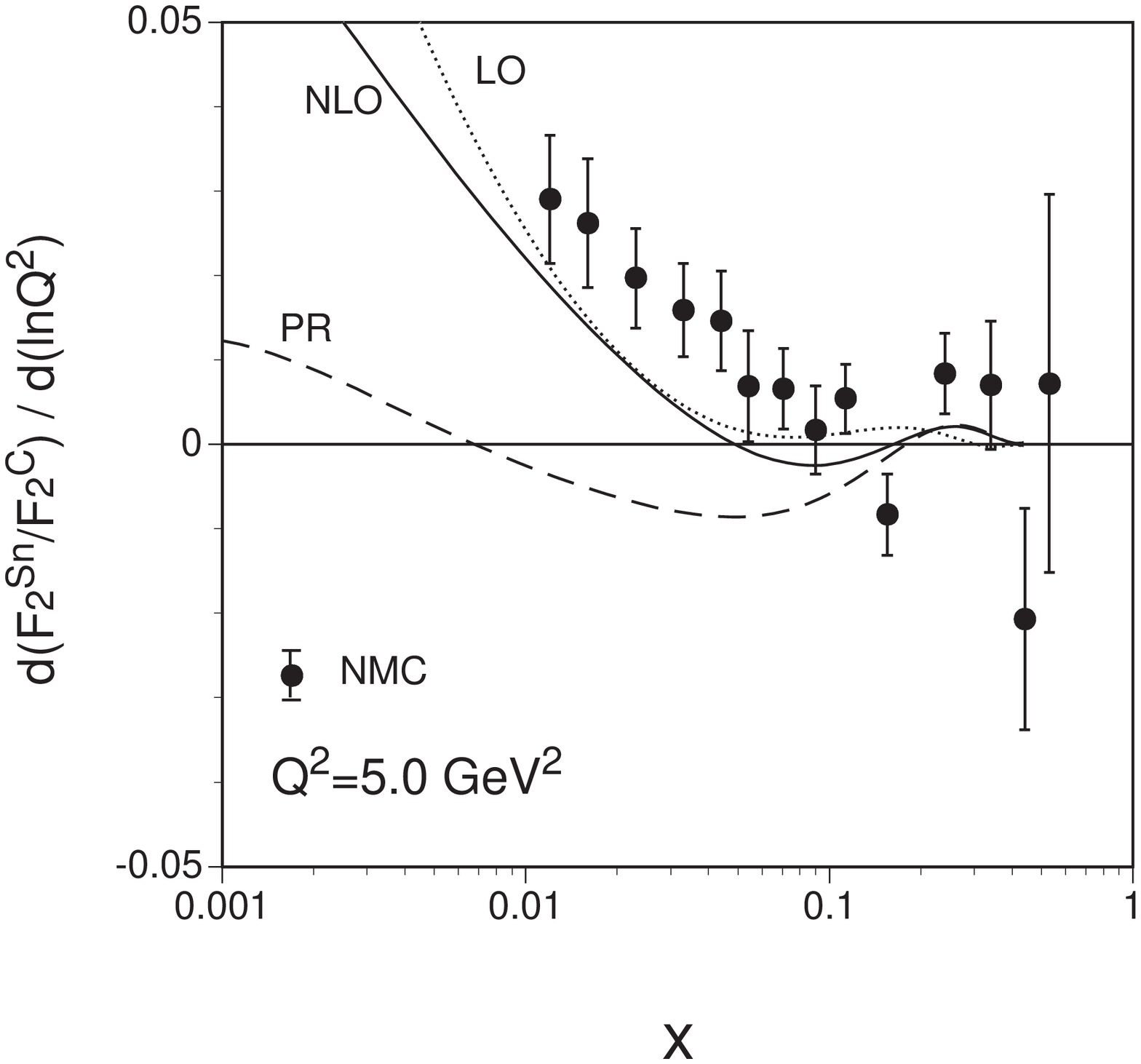,width=17cm}

%\vspace{3.0cm}
\centering{\Large{\bf Figure 1}}
%%%%%%%%%%%%%%%%%%%%%%%%%%%  figure 1 %%%%%%%%%%%%%%%%%%%%%%%%%%%%%%%%%%

\vfill\eject
%%%%%%%%%%%%%%%%%%%%%%%%%%%  figure 2 %%%%%%%%%%%%%%%%%%%%%%%%%%%%%%%%%%
\vspace*{-5.0cm}
\hspace*{-1.2cm}
\epsfile{file=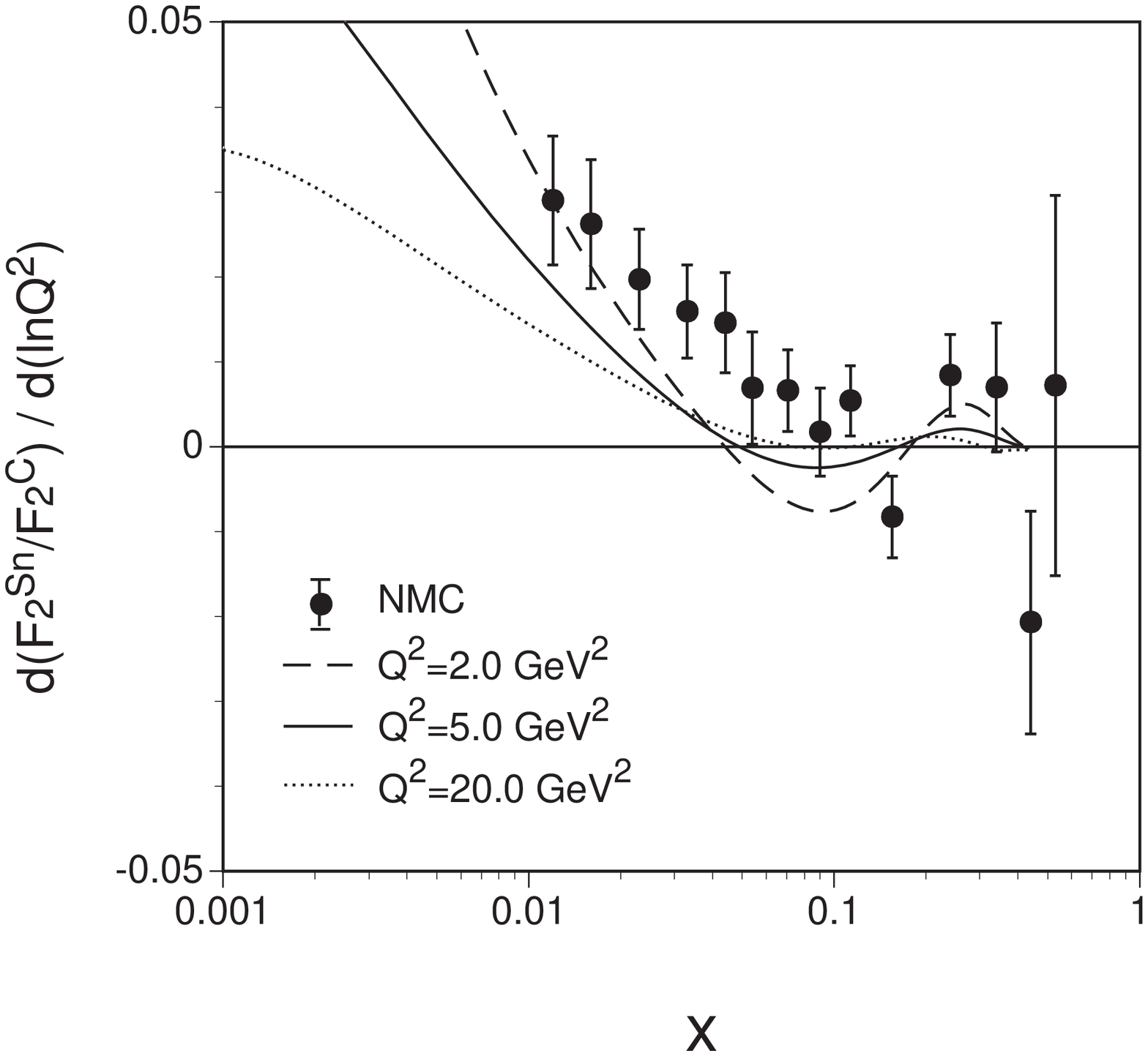,width=17cm}

%\vspace{3.0cm}
\centering{\Large{\bf Figure 2}}
%%%%%%%%%%%%%%%%%%%%%%%%%%%  figure 2 %%%%%%%%%%%%%%%%%%%%%%%%%%%%%%%%%%

\end{document}